\documentclass[12pt,a4paper]{article}
\usepackage{amsmath}
\usepackage[dvips]{graphicx}
\input epsf
\usepackage{times}

\def\be{\begin{equation}}
\def\ee{\end{equation}}
\def\bea{\begin{eqnarray}}
\def\eea{\end{eqnarray}}

\newcommand{\hup}{h^\uparrow}

\begin{document}
\begin{titlepage}
\title{The challenge of hyperon polarization}
\author{S. M. Troshin,
 N. E. Tyurin\\[1ex]
\small  \it Institute for High Energy Physics,\\
\small  \it Protvino, Moscow Region, 142280, Russia} \normalsize
\date{}
\maketitle
\begin{abstract}
We provide a brief topical outline of the
persisting  problem of hyperon polarization and consider some
future experimental prospects. Predictions
which deserve experimental verification are proposed.
 \\[2ex]
\end{abstract}
\end{titlepage}
\section*{Inroduction}
In the recent years a number of significant and unexpected spin
effects  were discovered.  They    demonstrated  that  the
spin degrees of freedom will be important even at TeV
energy scale. Currently spin experiments are conducted at
almost all existing  accelerators  and are planned  for
those under construction.    The experimental data on the
spin observables represent  a rather small fraction of the data on
particle interactions --- most of the data are related to
spin--averaged  ones. Despite that, these data prove to be
very important for understanding of particle interactions.
They provide information on the  spin dependence of the interactions,
which in some sense is  more fundamental than the dynamics studied
using spin--averaged observables. These data {\it clearly indicate that
high energy interaction dynamics  depends significantly
  on  the spin degrees of freedom}.
Last year with the start of the RHIC spin program
\cite{rhic}, spin studies are moving again to the forefront of high energy
physics.

The spin experiments aim at:
\begin{itemize}
\item
the study of the spin structure of the nucleon, i.e., how the proton's spin
state can be obtained from a superposition of Fock states with different
numbers of constituents with nonzero spin;
\item
the study of how the dynamics of constituent interactions depends on the
spin degrees of freedom;
\item
to understand chiral symmetry breaking and helicity non-conservation on the
quark and hadron levels;
\item
study the overall nucleon structure and long range dynamics.
\end{itemize}
Those issues are closely interrelated at the hadron level and
 the results of the experimental measurements
can be interpreted in terms of hadron  spin
structure convoluted with the constituent interaction dynamics.

In deep-inelastic scattering  data analysis in the framework
of perturbative QCD provides information on the longitudinal spin
 parton densities $\Delta q(x)$.
The study of the transverse spin parton  densities
$\delta q(x)$ has recently become also an equally actual  problem.
 However,
in the usual deep-inelastic  scattering the transverse
spin  contribute only as  higher--twist contributions.
The transverse spin densities can be studied in the
Drell-Yan processes
or in specific semi-inclusive DIS reactions \cite{rhic}.

In  soft hadronic interactions  significant single-spin
effects could be expected since the helicity conservation
does not work for interactions at large distances,
 once the chiral $SU(3)_L\times SU(3)_R$ symmetry of
 the QCD Lagrangian is spontaneously broken.
However, the
spin asymmetry $A_N$  at low transverse momentum
found to be small and decreases with energy.
In contrast, the spin asymmetry  increases at
high transverse momentum in elastic scattering,
where we should expect a decreasing behaviour
based on the helicity conservation due to the chiral invariance
of perturbative QCD. More theoretical work and experimental data will be
needed to understand the dynamics of these unexpected single-spin effects
observed in elastic scattering and inclusive hyperon and
meson production. In the following we consider these issues concentrating
on the particular problem of hyperon polarization.

\section{Experimental status}
In 1976 it  was
discovered    that in the reaction $pp\rightarrow \Lambda X$
  highly  polarized $\Lambda  $--hyperons  were  produced,  and    their
polarization  rises  with  $x_F$  and $p_{\perp }$ \cite{heller}.
 Nowadays there is
 a   rather   extensive  experimental set of data   on  the  hyperon
polarization in inclusive reactions \cite{newrev,bravar}. The fact
that polarization
of the $\Lambda $   enters in  the formula   for    the    proton    angular
distribution in the $\Lambda \rightarrow p\pi ^-$ decay, which does not
conserve  parity, allows to perform its  measurement.
 The polarization of $\Lambda $  is
obtained from the angular--distribution analysis of the decay products.
 After
 the  $\Lambda $--hyperon polarization discovery, the polarization of other hyperons
 was measured and they  also appeared to be  polarized.
 The experimental studies revealed the following regularities
\cite{newrev,bravar}:
\begin{itemize}
\item
In    proton--proton  interactions,  the  polarization  of  inclusively
produced $\Lambda $--hyperons is negative,  its direction is  opposite
  to  the
vector  $\vec{p}\times\vec{p}_\Lambda  $,  where  $\vec{p}$ is the incident
particle momentum. The polarization  rises  with  transverse  momentum  of
$\Lambda$--particle.    For   the   range   of   transverse   momentum
$p_{\perp} >0.8$ GeV/$c$ the dependence of  $P_\Lambda  $  on  the
transverse  momentum  becomes  weak.  In  this  region of $p_{\perp}$, the
polarization of  $\Lambda  $  grows  linearly  with  $x_F$.  The  $\Lambda
$--hyperon  polarization  has  been  measured  up to rather high values of
$p_{\perp}\simeq 3.5$ GeV/$c$.  However, even at the  maximum
values of $p_{\perp}$
 no tendency for a decreasing polarization  was observed.
 It is interesting that spin asymmetry $A_N$ and spin transfer parameter
 $D_{NN}$ have shown  similar $p_\perp$--dependence in $\Lambda$--production
 at FNAL \cite{bravar}.
  \item
 The $\Lambda $--hyperon polarization is energy independent  in
a wide  range of beam energy $p=12 - 2000$ GeV/c.
\item
In proton--proton interactions,  hyperon production reactions
reveal  the following relations between the hyperon polarizations:
\[
P_{\Sigma ^0}\simeq P_{\Sigma ^-}\simeq P_{\Sigma ^+}\simeq -P_{\Lambda }\simeq
-P_{\Xi ^0}\simeq -P_{\Xi ^-},
\]
\[
P_{\bar{\Lambda }}\simeq P_{p}\simeq P_{\Omega ^-}\simeq 0,\quad
P_{\Lambda }(p\rightarrow \Lambda )\simeq P_{\bar{\Lambda }}(\bar{p}\rightarrow
\bar{\Lambda }).
\]
\item
 For the reactions involving mesons the following relations are observed:
\[
P_\Lambda (K^-\rightarrow \Lambda )>0,\quad
|P_\Lambda    (K^-\rightarrow    \Lambda    )|\gg   |P_\Lambda
(p\rightarrow \Lambda )|
\]
\end{itemize}
and
\[
|P_{\bar{\Lambda }}(K^+\rightarrow \bar{\Lambda })|\gg
|P_{\bar{\Lambda }}(\bar{p}\rightarrow \bar{\Lambda })|.
\]
 The polarization of $\Xi ^-$--hyperons produced
by 800--GeV protons has also been measured at FNAL. Contrary to the
regularities  observed in
the $\Lambda $--production,  energy dependence of
$P_{\Xi ^-}$ has been observed for the first time. The quantity
 $P_{\Xi ^-}$ also
does not show the $x_F$--dependence as observed in
$\Lambda $--polarization data.

To determine the polarization of  directly produced hyperons
  one has to measure the hyperon polarization  in  exclusive
reactions. To this  end,  the  $\Lambda $--hyperon  polarization
has been studied
\cite{dif} in  the exclusive diffractive process
\[
p+p\rightarrow p+[\Lambda ^0K^+].
\]
Results of the measurements demonstrated  that
$\Lambda $--hyperon polarization increases linearly with
$p_{\perp}$ at $\sqrt{s}\simeq  30$ GeV and gets a very large value,
$P_\Lambda \simeq 80\%$, at $p_{\perp}\simeq 1$ GeV/c.

Therefore, since the $\Lambda $ polarization does not show a
decrease at large $p_{\perp}$ values, the data on hyperon
polarization  provide a direct evidence for the $s$--channel helicity
nonconservation in hadron collisions in a wide energy and $p_{\perp}$
range.
\section{Theoretical status}
The theoretical status of hyperon polarization was described in many review papers
(see e. g. \cite{rev}). Hyperon polarization  even the most simple
case --- the polarization of $\Lambda$'s --- is not understood in
 pQCD and it seems that in the future it could become an
even more serious problem than the nucleon spin problem.
Those problems are likely related.  One could
attempt to connect the spin structure  studied in
deep--inelastic
scattering with the polarization of $\Lambda$'s observed in
hadron production \cite{pollam}.
As is widely known now, only a part (less
than one third) of the proton spin is due to spins of the quarks.
Studies of spin effects in inclusive processes  yield
information on the contribution of the spin and orbital
angular momenta of  quarks and gluons
into the  hadron helicity:
\begin{equation}
1/2=1/2\Delta\Sigma+L_q+\Gamma+L_g,
\end{equation}
In the  sum above  first two terms are the contributions of spin and orbital
momentum of the quarks and last two terms are the corresponding
quantities related to the gluons, i.e.
all these terms have a clear physical interpretation;
however  besides the first one, these are gauge
and frame dependent.

Experimental results on the nucleon structure  can be interpreted in an
effective QCD approach ascribing a substantial part of the hadron spin
to  orbital angular momentum of the quark matter.  It is natural to
assume that this orbital angular momentum could be the origin of
the asymmetries in hadron production.
It is also evident from the recent deep--inelastic scattering data
that
strange quarks play essential role in the  proton structure and in
its spin balance.
Polarization effects in  hyperon production are complimentary
to these results and demonstrate \cite{newrev} also that strange
quarks acquire polarization in the course of the hadron interactions.
The explicit formula
for the asymmetry in  terms of the helicity amplitudes has the
following form:
\[
P,\;A_N=2\frac{\sum_{X,\lambda _X, \lambda _2}\int d\Gamma _X
\mbox{Im}[F_{\lambda _X;+,\lambda _2}F^*_{\lambda_X;-,\lambda
_2}]} {\sum_{X,\lambda _X;\lambda _1 \lambda _2}\int d\Gamma _X
|F_{\lambda_X;\lambda _1 ,\lambda _2}|^2},
 \]
and shows that non-zero single spin asymmetries require the presence
of  helicity flip and nonflip amplitudes and a phase difference
between those amplitudes.
A straightforward application of  perturbative QCD using collinear
factorization scheme, where e. g.
\begin{equation}
A_N d\sigma\sim \sum_{ab\rightarrow cd}\int
d\xi_Ad\xi_B\frac{dz}{z}\delta f_{a/A}(\xi_A) f_{b/B}(\xi_B)\hat
a_N d\hat \sigma _{ab\rightarrow cd} D_{C/c}(z) \end{equation}
encounters serious difficulties in trying to explain the
measured polarization results. This is due to the
chiral invariance and vector nature of QCD Lagrangian  which
provide important consequences for spin observables and therefore allows
for direct and unambiguous experimental test of perturbative QCD.
 Indeed, the QCD Lagrangian is invariant   under
transformation of $SU(3)_L\times SU(3)_R$ group ($N_f=3$) and therefore
QCD interactions are the same for  left and right quarks:
\[
\bar{\psi }\gamma _\mu \psi A^\mu =\bar{\psi }_L\gamma _\mu \psi _LA^\mu
 +\bar{\psi }_R\gamma _\mu \psi _RA^\mu \,.
\]
 Left--handed  (right--handed) massless
particles will always
stay left handed  (right--handed).
For massless quarks chirality and helicity coincide.
If the quarks have a non--zero mass  chirality and helicity
 are only  approximately equal at
high energies, namely:
\[
\psi _{1/2}=\psi _R+O\left(\frac{m}{\sqrt{\hat{s}}}\right)\psi _L,
\quad
\psi _{-1/2}=\psi _L+O\left(\frac{m}{\sqrt{\hat{s}}}\right)\psi _R\,
\]
where $\pm 1/2$ are the quark helicities.
Any quark line entering a Feynman diagram
corresponding the QCD
Lagrangian  emerges from it
with unchanged helicity.
 Quark helicity conservation is the most characteristic feature of
this theory.
To get quark polarization  values $\hat{P}_q\neq 0$ it is necessary that
helicity flip amplitude is a  non--zero  and
that phases of helicity flip $F_f$ and non--flip
$F_{nf}$ amplitudes are different, since
\[
\hat{a}_N,\; \hat{P}_q\propto \mbox{Im}(F_{nf}F_f^*).
\]
In QCD the quark helicity flip amplitude is of order of
$ m/\sqrt{\hat{s}}$. In the Born
approximation the amplitudes are real and therefore
one needs to consider the loop diagrams , i.e. the
quark helicity flip amplitude will be proportional to
\[
F^q_f\propto \frac{\alpha _sm_q}{\sqrt{\hat{s}}}F_{nf}^q,
\]
 and consequently the quark polarization is vanishingly small in hard
interactions \cite{pump}:
\[
\hat{a}_N,\;\hat {P}_q\propto \frac{\alpha _sm_q}{\sqrt{\hat{s}}}
\]
due to  large value of the hard scale $\sqrt{\hat{s}}\sim
p_{\perp}$ and small values of  $\alpha _s$ and $m_q$. Here $m_q$
stands for  the mass of the current quark.
 Explicit
calculation of $s$-quark polarization in gluon fusion process
yield  $\hat{P}_s\leq 4$\% \cite{gold}.
 Thus, a {\it straightforward collinear factorization leads
to very small $P_\Lambda $}.

Several modifications of this simple perturbative QCD scheme have been
proposed. The factorization formula
including  higher twist
 contributions was obtained in \cite{sterm} and it was recently
  proposed to consider
 also  twist-3 contributions \cite{efrem}
\[
{ E_{a/A}(\xi_{1A},\xi_{2A})}\otimes q_{b/B}(\xi_B)\otimes \delta D_{c/\Lambda}
(z)\otimes d\hat{\sigma}_{ab\to cd}
\]
as a source for the strange
quark polarization \cite{koike}.
Here the function $E_{a/A}(\xi_1,\xi_2)$ is a
 twist-3 unpolarized distribution, and higher twist partonic correlations could
also give additional contributions to fragmentation functions.
Since  there is no  information
  on higher
 twist contributions, additional
assumptions and models are necessary. Moreover it is not clear what  the
 dynamical origin is of strange
quark polarization that arising through the unpolarized quark-gluon correlator
$E_{a/A}(\xi_1,\xi_2)$.
{\it The predicted decreasing dependence
  $P_\Lambda\sim 1/p_\perp$ at high $p_\perp$
still does not correspond to the experimental trends}.

Note that the situation   now appeares to be complicated
  even for the standard
leading twist parton distributions
 which,  are not simply related to the structure functions
 measured in deep-inelastic scattering as it was commonly accepted until now
 \cite{brod}.
The crucial role here belongs  to the final-state interactions, which could provide
 a phase difference and lead to   non-zero single-spin asymmetries in
 deep-inelastic scattering \cite{brodas}.

 Another
modification of the collinear factorization takes into account a
role of $k_\perp$-effects.
By analogy with Collins' suggestion for the fragmentation of a transversely
polarized quark\footnote{It should be noted, that the contributions of  instantons
 to the quark fragmentation
could provide necessary helicity flip, but it also leads to  vanishingly small
 single spin asymmetry at large $p_\perp$ starting with $p_\perp\simeq 2.5$ GeV/c
  \cite{koch}.},
 it was proposed to consider  so called polarizing fragmentation
functions (as it was demonstrated recently by Collins \cite{coll},
the $k_\perp$-effects
 in distribution functions \cite{sivers} proposed by Sivers,
 are, in fact, compatible
  with $T$--invariance, but those are not relevant for
   the case of the hyperon polarization in the
  processes where initial hadrons are unpolarized)
for an unpolarized quark with momentum $\bf p_q$ to fragment into
a spin 1/2 hadron $h$ with momentum ${\bf p_h} = z {\bf p_q} + \bf k_\perp$
and polarization vector ${\bf P}_h$  \cite{anselm}:
\[
 D_{\hup/q}(z, {\bf k_\perp}) = \frac 12 \>  D_{h/q}(z, k_\perp)
 +
\frac 12 \> \Delta^ND_{\hup/q}(z, k_\perp) \>
\frac{\hat{{\bf P}}_h \cdot ({\bf p_q} \times {\bf k_\perp)}}
{|{\bf p_q} \times {\bf k_\perp}|}
\]
in the framework of the generalized
factorization scheme -- with the inclusion of intrinsic transverse
motion -- along with pQCD dynamics.  Thus, in this approach,
 the source of $\Lambda$
 polarization was shifted into the
 polarizing fragmentation function.
Some features of this scheme are:
\begin{itemize}
  \item a falling $\sim \langle k_\perp \rangle/p_\perp$
   dependence of the polarization at large $p_\perp$,
  \item still need a dynamical model for polarizing fragmentation functions,
  \item it is not compatible with the $e^+e^-$ - data at LEP, where
  no significant transverse
  polarization of the $\Lambda$ was found  \cite{aleph}.
\end{itemize}
{\it The overall conclusion  is that the dynamics of $\Lambda$
 polarization as well as of
other single spin asymmetries within pQCD (leading twist collinear factorization
or its modifications) is far from being settled}.
The essential point here is the assumption
 that at  short distances the vacuum is perturbative.
 The study of the $p_{\perp}$--dependence of the one--spin
asymmetries can be used as a way to reveal the transition from the
non--pertur\-ba\-tive phase ($P\neq 0$) to the perturbative
phase ($P=0$) of QCD. The very existence of such transition can
 not be taken
for granted since the vacuum, even at
short distances, could be filled with the fluctuations of gluon or quark
fields.
The measurements of the one--spin transverse
asymmetries and polarization  will be an important probe of the
 chiral structure of the effective QCD Lagrangian.

At the same time we can note that polarization effects
as well as some other recent
experimental data demonstrate
that hadron interactions have a significant degree of coherence.
The persistence of elastic scattering at  high energies and significant
polarization effects  at large angles  confirm
 this \cite{plus}.
It means that the polarization dynamics has its roots hidden in the
genuine non-perturbative sector of QCD. Several models exploiting confinement
and chiral symmetry breaking have been proposed. The best known  are
the model based on the classical string picture (Lund model)
where the polarization of a strange
quark is the result of angular momentum conservation and the model based on the
polarization of strange quark due to the Thomas precession.
 The main feature of the both models
is that these take into account for the consequences of  quark confinement.
 The most recent
discussion of the successes and failures of these models can be found in
\cite{rev}.

It is worth noting  that the chiral  $SU(3)_L\times SU(3)_R$
group of the QCD Lagrangian is not observed in the hadron spectrum.
 In  the real world this group is broken down to $SU(3)_V$
with  the appearance of the $N_f^2-1=8$ Goldstone bosons ($\pi $, K and
$\eta $).
Chirality, being a symmetry of QCD Lagrangian, is broken (hidden) by the vacuum
state
\[
\langle 0|\bar{\psi }\psi |0 \rangle=
\langle 0|\bar{\psi }_L\psi _R+\bar{\psi }_R\psi _L|0
\rangle.
\]
The scale of the spontaneous chiral symmetry breaking
is $\Lambda_\chi\simeq4\pi f_\pi\simeq 1$ GeV, where $f_\pi$ is the pion
decay constant.
Chiral symmetry breaking generates  quark masses:
\[
 m_U = m_u-2g_4\langle 0|\bar u u|0\rangle-2g_6\langle 0|\bar
d d|0\rangle \langle 0|\bar s s|0\rangle .
\]
Massive  quarks are quasiparticles (current quarks
surrounded by  cloud of quark--antiquark pairs of different flavors).
It is worth to stress that in addition to $u$
 and $d$ quarks the constituent quarks ($U$, for example) contain pairs
of strange quarks  and the ratio of the scalar
density matrix elements \begin{equation} y=
{\langle U| \bar ss|U\rangle}
/ {\langle U|\bar u u+\bar d d+\bar s s|U\rangle}
 \label{str} \end{equation} is estimated  as $y=0.1 - 0.5$.
The spin of the constituent quark $J_{U}$  is
  \[
 J_{U}=1/2  =  S_{u_v}+S_{\{\bar q q\}}+ L_{\{\bar qq\}}=
               1/2+S_{\{\bar q q\}}+ L_{\{\bar qq\}}.
\]
\[
(\Delta\Sigma)_U=1/2+S_{\{\bar q q\}},\quad
(\Delta\Sigma)_p = (\Delta U+\Delta
D) (\Delta\Sigma)_U,\quad
(\Delta\Sigma)_p\simeq 0.2,
\]
leading to $ L_{\{\bar q q\}}\simeq 0.4$
Accounting for the
axial anomaly in the framework of chiral quark models results in
the compensation of the valence quark helicity by  helicities of quarks
 from the cloud in the structure of constituent quark. The specific
 nonperturbative mechanism of such compensation may be different in
 different models.

Orbital angular momentum, i.e.
 orbital motion of quark matter inside constituent quark, could
 generate  the observed asymmetries in inclusive production on a polarized
 target at
  moderate and high transverse momenta \cite{asm}.
The main points of the mechanism proposed in \cite{asm} are:
\begin{itemize} \item the asymmetry in the pion production
reflects the internal spin structure
of the constituent quarks, i.e. it arises from the orbital angular
momentum of the current quarks inside the constituent quark structure;
\item the sign
of the asymmetry and its value are proportional to the polarization of the
 constituent quark inside the polarized nucleon.
\end{itemize}
Another model \cite{berl} which relates the asymmetries to the
  orbital motion considers these  a result of the
 orbiting valence quarks. It is not yet clear if
  it is possible to reconcile the orbital motion of valence
 quarks with  the spectroscopy data.

The generic behavior of the different asymmetries in inclusive
 meson and hyperon production
\begin{equation}
P,\; A_N(s,x,p_{\perp})=\frac{ \sin[{\cal{P}}_{\tilde Q}(x)
\langle L_{\{\bar q q\}}\rangle] {W_+^h(s,x,p_\perp)}}{ {
W_+^s (s,x,p_\perp)+W_+^h(s,x,p_\perp)}},
 \end{equation}
(where the  functions $
W_+^{s,h}$ are determined by the interactions at large and small
distances) was predicted \cite{asm} to have a characteristic
$p_{\perp}$--dependence, in particular: vanishing asymmetry for
$p_{\perp}<\Lambda_\chi $, an increase in the region of
$p_{\perp}\simeq\Lambda_\chi $, and a $p_{\perp}$--independent (flat)
asymmetry for $p_{\perp}>\Lambda_\chi $ (Fig. 1).
\begin{center}
\begin{figure}[h]

\vspace{2mm}

\begin{center}
\epsfxsize=2.1in \epsfysize= 1.5in
 \epsffile{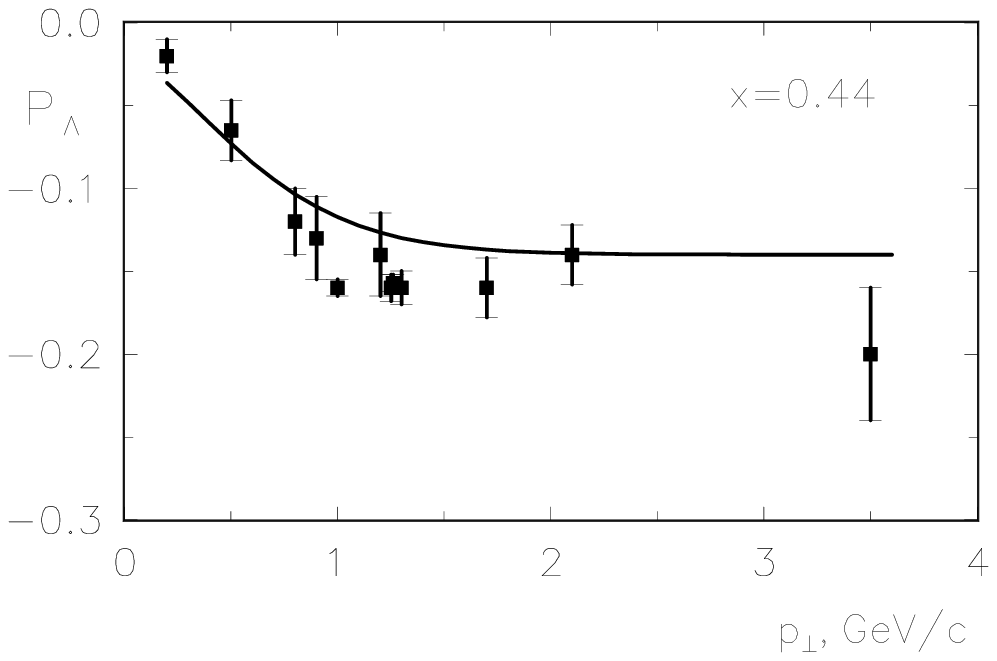}\quad
\epsfxsize=1.8in \epsfysize=2.1in
 \epsffile{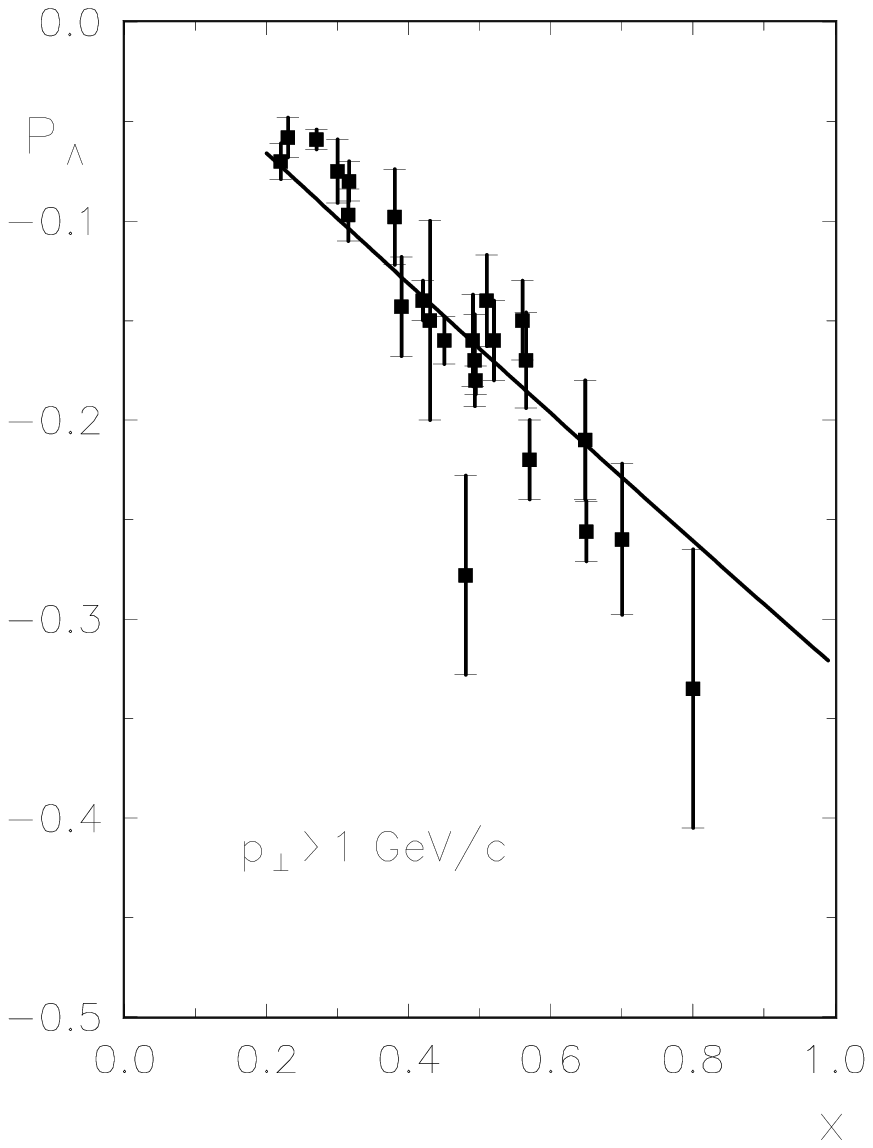}
\end{center}
 \caption[pdep]{Transverse momentum and Feynman $x$ dependence of $P_\Lambda$.}
\label{fig:1}
\end{figure}
\end{center}
The parameter
$\Lambda_\chi\simeq 1$ GeV/c is determined by the scale of spontaneous
chiral symmetry  breaking.  This behavior of the asymmetry  follows
from the fact that the constituent quarks themselves have a slow (if at
all) orbital motion and are in an $S$--state.  Interactions with
$p_{\perp}>\Lambda_\chi $ resolve the internal structure of
constituent quark and ``feel'' the presence of internal orbital momenta
inside this constituent quark.
The polarization of $\Lambda$--hyperons in the model \cite{pollam}
 is  result of the polarization
the constituent quark due to multiple scattering in the effective
field.
It means that $s$-quark in the structure of constituent quark will be
polarized too, since in the model the spins of
quark--antiquark pairs  and their angular orbital momenta are exactly
compensated:
\begin{equation}
 L_{\{\bar q q\}}= -S_{\{\bar q q\}}.\label{corr}
\end{equation}
It should be noted  that DIS data show that strange quarks are
 negatively polarized in polarized
nucleon, $\Delta s\simeq -0.1$.
Also elastic $\nu p$-scattering data  provide a negative value
 $\Delta s=-0.15\pm 0.08$ \cite{nu}.
The presence and polarization of strange quarks inside a hadron should
also give an experimental signal in hadronic reactions.

The parent constituent quark and the strange quark
have opposite polarization.
The size of the constituent quark is determined by the scale of
the chiral
symmetry breaking, i. e. $R_Q\simeq 1/\Lambda_\chi$.
The polarization is
significant when the interaction resolves the internal structure
of the constituent quark: $r<R_Q$
i. e. for $p_{\perp}>\Lambda_\chi\simeq 1$ GeV/c.
It is useful to note here that  current quarks appear in
the nonperturbative vacuum and become
quasiparticles due to the nonperturbative ``dressing up''
 with a cloud of $\bar q q$-pairs. The mechanism of this process
  could be associated with the
strong coupling existing in the pseudoscalar channel.

{\it Thus, this model \cite{pollam} predicts a
 similarity of the $p_\perp$--dependencies for
the different spin observables: increase up to $p_{\perp}=\Lambda_\chi$
and flat dependence  at  $p_{\perp}>\Lambda_\chi$}.

The transition to the partonic picture
 in this model is described by the introduction of a momentum cutoff
 $\Lambda=\Lambda_\chi\simeq 1$ GeV/c, which corresponds to the scale
of spontaneous chiral symmetry  breaking.
We shall  see that at higher $p_\perp$
 the constituent quark to be a cluster of partons,
 which however should  preserve  their orbital
 momenta, i.e. the
 orbital angular momentum will be retained
 and the partons in the cluster are correlated.
Note, that the short--distance interaction in this
approach observes a coherent rotation of correlated  $\bar q q$--pairs
inside the constituent quark and not a gas of the free  partons.
The nonzero internal orbital
momenta of partons in the constituent quark means that there are
significant multiparton correlations.
Indeed, a high locality of strange sea in the nucleon was found experimentally
by the CCFR collaboration at FNAL. This locality serves as a measure of the
local proximity of strange quark and
antiquark in momentum and coordinate space. It was shown \cite{ji} that the
CCFR data indicate that the strange quark and antiquark have very similar
distributions in  momentum and coordinate space.

\section{Experimental prospects}
Experimentally observed persistence and constancy of $\Lambda$--hyperon
polarization
 means  \cite{pollam}
 that chiral symmetry is not restored in the region
 of energy and values of $p_{\perp}$ where experimental measurements
 were performed. Otherwise we would not
 have any constituent quarks and should expect
 a vanishing polarization of $\Lambda$ \cite{pollam}. It is interesting
 to perform  {\it $\Lambda$--polarization measurements at RHIC and the LHC.
 It would allow to make a direct check of perturbative QCD and allow to
 make a cross-check of the QCD background estimations based on
 perturbative calculations for the LHC}. It is an important problem which
 still awaits its solution.

 Experimentally measurable effects of 10\%
  or higher are expected at c.m.s. energy
  $\sqrt{s}=500$ GeV for values of pseudorapidity
  $|\eta|\geq 2.3$, and for energies 2 TeV and 10 TeV
  at $|\eta|\geq 3.7$ and $|\eta|\geq 5$ respectively.

 The use of
a polarized beam will allow  to measure the two--spin
correlations in  hyperon production at TeV energies at the LHC, and to
reveal the  underlying mechanism.
 Moreover, measurements of the transverse and longitudinal asymmetries
provide information on  production mechanism and on
hadron spin structure. It seems very promising  to study
reactions with weakly decaying baryons in the final state such as
\begin{equation}
p_{\uparrow,\rightarrow}+p\rightarrow \Lambda _{\uparrow,\rightarrow}+X
\end{equation}
and to measure the parameters $D_{NN}$ and $D_{LL}$.
These asymmetries  are calculable in the
framework of perturbative QCD \cite{craig,bsrt}.
The data on
 $D_{NN}$ and $D_{LL}$ in the fragmentation region at large
$x_F$ seem to be interesting also from the
point of view of the polarization of the strange sea and strangeness
content of the nucleon. The strange sea
has a large negative polarization   value  according to the
interpretation of the polarized deep--inelastic scattering
results,  and it should be revealed in $D_{NN}$ and $D_{LL}$
behavior.
The typical behavior of spin parameters can be illustrated by the
dependence of asymmetry $A_N$ on $p_\perp$ (Fig. 2)
\begin{center}
\begin{figure}[t]
\vspace{2mm}
\begin{center}
\epsfxsize=2in \epsfysize=2in
 \epsffile{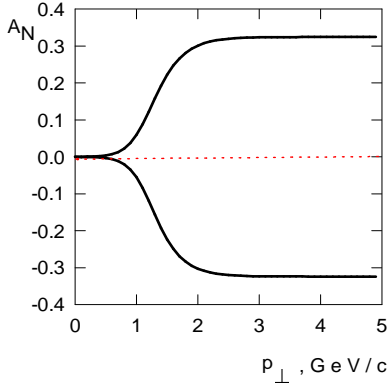}
\end{center}
 \caption[andep]{{The dependence of $A_N$ in charged pion production
 at $\sqrt{s}=500$ GeV and $x=0.5$ }}
\label{fig:3}
\end{figure}
\end{center}
The asymmetries have energy independent values about
30\% at
$p_\perp>\Lambda_\chi
(\simeq 1-2$ GeV/c) and $x\simeq 0.5$.

Note, that the corresponding asymmetry in  neutral pion
production has  a value about $5-6\%$ in the same kinematical
region.

If  two polarized nucleons are available in the  initial  state
then one could measure
three--spin  parameters
such as $(n,n,n,0)$ and $(l,l,l,0)$ in the process:
\begin{equation}
p_{\uparrow,\rightarrow}+p_{\uparrow,\rightarrow}=\Lambda _{\uparrow,\rightarrow}+X,
\end{equation}
where the polarization of $\Lambda $--hyperon is studied through its
decay.
Measurements of the three--spin correlation parameters would provide
important data for the study of  hyperon production dynamics and
mechanisms for hyperon polarization. On that basis
we expect, as it was already mentioned, similar $p_\perp$--dependence
for all spin parameters when  initial hadrons have no valence strange
quarks.

When  one of the colliding hadrons has a valence strange
quark, the picture will include  two mechanisms: the one described above
and another one -- polarization of valence strange
quarks in the effective field. The latter mechanism will provide
 $\Lambda$'s polarized in the opposite direction and can explain
 the positive
 large polarization $P_\Lambda$ in $K^-p$--interactions.

As it was claimed in \cite{vect} on the basis of DIS nonperturbative
analysis, that the orbital angular momentum increases with virtuality $Q^2$. Thus,
in the framework of the model studied and according to the
relation (\ref{corr})
it would lead to increasing  polarization of strange quark.
It means, in particular, that the transverse polarization
$P_\Lambda(Q^2)$ in the current
 fragmentation region of the semi--inclusive process
  \[ l+p\to l+\Lambda_\uparrow+ X \] will
also increase with virtuality, provided  the competing mechanism
 of the polarization of the valence strange quark from the $\varphi$--meson
 gives a small contribution. Regardless of  these particular predictions,
 it seems that the experimental studies of  polarization dependence
 on virtuality could shed  new light on this problem.

\section{Quark--gluon plasma detection}
On the base of the model in \cite{pollam}  one  expects
 zero polarization in the
region where QGP  has  formed, since  chiral symmetry is restored
and there is no room for  quasiparticles such as  constituent quarks.
The absence or strong diminishing of transverse  hyperon polarization
can be used
therefore as a signal of QGP formation in heavy-ion collisions.
This prediction should also be valid
for the  models based on confinement, e.g. the Lund and Thomas precession
model. In particular, the polarization of $\Lambda$ in heavy--ion collisions
in the model based on the Thomas precession was described in  \cite{ayal}
where nuclear effects were discussed as well.
However, we do not expect a strong diminishing of the $\Lambda$--polarization due to
the nuclear effects:  the available data show a weak $A$--dependence and are
not sensitive to the type of the target. Thus, we could use a vanishing polarization
of $\Lambda$--hyperons in heavy ion collisions as a sole result of QGP formation provided
the corresponding observable is non-zero  in  proton--proton collisions.
 The prediction based on this observation
would be a decreasing behavior of polarization of $\Lambda$ with the impact parameter
in heavy-ion collisions in the region of energies and densities where QGP was
produced:
\begin{equation}\label{zer}
P_\Lambda(b)\to 0\quad \mbox{at}\quad b\to 0,
\end{equation}
since the overlap is maximal at $b=0$. The value of the impact parameter  can be
controlled by the centrality in heavy--ion collisions.
The experimental program should therefore
include  measurements of $\Lambda$--polarization
in $pp$--interactions first, and then if a significant polarization would be
measured,  the corresponding measurements could be a useful tool for the
 QGP detection.
Such measurements seem to be experimentally feasible
at  RHIC and  LHC provided it is supplemented with forward
detectors.

\section*{Conclusion}
In conclusion we would like to emphasize the main points of our
discussion:
\begin{itemize}
\item
a universal $p_\perp$--dependence for all spin parameters
in $\Lambda$--hyperon production which reflects
a finite size for constituent quarks is predicted: all spin parameters in the
collisions of hadrons without valence strange quarks demonstrate an increase
in absolute value up to $p_\perp\simeq 1$ GeV/c and then become flat.
\item
three--spin correlation parameters -- the new observables which can be measured
in hyperon production with polarized beams -- can provide  new insight
into the mechanism of hyperon polarization
\item
hyperon polarization should vanish in heavy--ion collisions
when quark--gluon plasma is formed, its
decrease with centrality
can be considered as a gold--plated signature of QGP formation.
\end{itemize}

\section*{Acknowledgements}
We are grateful  to  A. De Roeck for the useful suggestions, comments
 and careful reading of the manuscript,  V. Petrov and P. Schlein for
the interesting discussions.

\end{document}